\documentclass[pra,twocolumn,showpacs]{revtex4-1}     
\usepackage{epsfig,amsmath} 

\begin{document}

\title{Phase boundary of spin-polarized-current state of electrons in bilayer graphene} 

\author{Xin-Zhong Yan$^{1}$, Yinfeng Ma$^{1}$, and C. S. Ting$^2$}
\affiliation{$^{1}$Institute of Physics, Chinese Academy of Sciences, P.O. Box 603, 
Beijing 100190, China\\
$^{2}$Texas Center for Superconductivity, University of Houston, Houston, Texas 77204, USA}

\date{\today}

\begin{abstract}
Using a four-band Hamiltonian, we study the phase boundary of spin-polarized-current state (SPCS) of interacting electrons in bilayer graphene. The model of spin-polarized-current state has previously been shown to resolve a number of experimental puzzles in bilayer graphene. The phase boundaries of the SPCS with and without the external voltage between the two layers are obtained in this work. An unusual phase boundary where there are two transition temperatures for a given carrier concentration is found at finite external voltage. The physics of this phenomenon is explained.
\end{abstract}

 
\maketitle

\section{Introduction}

From a framework of free-electron system in bilayer graphene (BLG), there can be a tunable gap between the conduction and valence bands under an external electric field. Because of this property, BLG is a promising material with a great potential for application to new electronic devices \cite{Ohta,Oostinga,McCann,Castro}. The experimental observations on high quality suspended BLG samples \cite{Weitz,Freitag,Velasco,Bao} has revealed that the ground-state of the electron system at the charge neutrality point (CNP) is insulating with a gap about 2 meV that can be closed by a perpendicular electric field of either polarity. In an external magnetic field, the gap grows greatly with increasing the magnetic field much larger than the Zeeman splitting \cite{Velasco}. The observed quantum-Hall states at the integer fillings from $\nu$ = 0 to $\pm 4$ \cite{Elferen,Velasco1} are different from the prediction of free-electron model by which the $\nu$ = 0 state should be eightfold degenerated. These puzzling properties of the system at low temperature stem from the electron interactions. A number of theoretical models for the ground state of the interacting electron system in BLG has been proposed \cite{Min,Nandkishore,Zhang1,Jung,Nilsson,Gorbar,Zhang2,Milovanovic,Zhu,Yan1,Yan2,Yan3,san-jose,Lemonik}. Among these theories, the experimental observations can be reasonably explained only by the model of spin-polarized current state (SPCS) for the electrons \cite{Yan3}. The SPCS is a symmetry-broken state due to the electron interactions at low temperature and at low carrier concentration. For application of BLG, it is necessary to know the phase boundary of the SPCS. 

In this work, we intend to investigate the phase transition between the SPCS and the normal state of the interacting electrons in the BLG with and without external voltage between the two layers. Using the four-band model for the electrons, we derive and solve the equation for the phase boundary of the SPCS. At finite voltage, the electron system can be in a state with the layer-charge polarization (LCP). Above the LCP background, there may exist spin-polarized-current ordering. The phase transition between the SPCS with a LCP background and the state of the pure LCP should be unusual. This study not only is of the scientific interest but also provides the knowledge for real application of the BLG.

\section{Spin-polarized-current state}

The lattice of the BLG shown in Fig. 1 (left) contains atoms $a$ and $b$ on top layer, and $a'$ and $b'$ on bottom layer with lattice constant $a \approx 2.4$ \AA~ and interlayer distance $d \approx 3.34$ \AA. The Hamiltonian of the electron system in BLG is
\begin{eqnarray}
H&=&-\sum_{ij\sigma}t_{ij}c^{\dagger}_{i\sigma}c_{j\sigma} +U\sum_{j}\delta n_{j\uparrow}\delta n_{j\downarrow} +\frac{1}{2}\sum_{i\neq j}v_{ij}\delta n_{i}\delta n_{j}\nonumber\\
\label {hm}
\end{eqnarray}
where $c^{\dagger}_{i\sigma} (c_{i\sigma})$ creates (annihilates) an electron of spin $\sigma$ at site $i$, $t_{ij}$ is the hopping energy between sites $i$ and $j$, $\delta n_{i}=n_i-n$ is the number deviation of electrons at site $i$ from the average occupation $n$, and $U$ and $v$'s are the Coulomb interactions between electrons. By the tight-binding model, we consider only the intra-layer nearest-neighbor (NN) [between $a$ ($a'$) and $b$ ($b'$)] electron hopping with $t$ = 3 eV and inter-layer NN (between $b$ and $a'$) electron hopping with $t_1$ = 0.273 eV \cite{Tatar,LMZ}. 

We use the mean-field theory (or the self-consistent Hartree-Fock approximation) (MFT) to treat the interactions. By the MFT, the interaction part in Eq. (\ref{hm}) is approximated as
\begin{eqnarray}
H_{int}&=&U\sum_{j\sigma}\langle\delta n_{j\bar\sigma}\rangle\delta n_{j\sigma} +\sum_{i\neq j}v_{ij}\langle\delta n_{i}\rangle\delta n_{j}  \nonumber\\
& &+\sum_{i\neq j,\sigma}v_{ij}\langle c_{i\sigma}c^{\dagger}_{j\sigma}\rangle c^{\dagger}_{i\sigma}c_{j\sigma},  \label{mft}
\end{eqnarray}
where the first and second lines in the right-hand side of Eq. (\ref{mft}) are respectively the Hartree and Fock factorizations, and $\bar\sigma$ means the inverse spin of spin $\sigma$. According to the many-particle theory, while the direct interactions in the Hartree term are given by the bare Coulomb interactions, the interactions in the exchange part include the screening due to the electronic charge fluctuations. We will 
adopt effective exchange interactions \cite{Yan3,Hwang} that qualitatively take into account the screening effect. From Eq. (\ref{mft}), we extract out the self-energy of the spin-$\sigma$ electron,  
\begin{eqnarray}
\Sigma^{\sigma}(i,j)&=&(U\langle\delta n_{j\bar\sigma}\rangle +\sum_{j'\ne j}v_{j'j}\langle\delta n_{j'}\rangle)\delta_{ij}  \nonumber\\
& &+v^{eff}_{ij}(\langle c_{i\sigma}c^{\dagger}_{j\sigma}\rangle-\langle c^{\dagger}_{j\sigma}c_{i\sigma}\rangle)/2|_{i\neq j},  \label{ser}
\end{eqnarray}
where $v^{eff}$ means the effective interactions with electron screenings. 

\begin{figure}[t]
\centerline{\epsfig{file=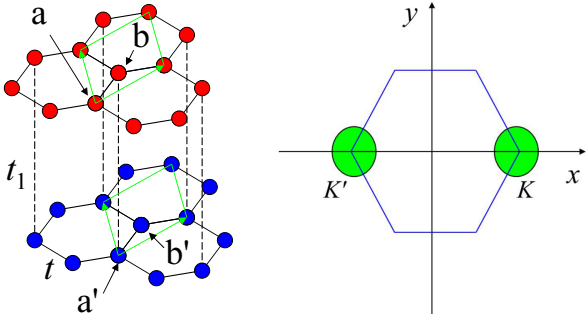,width=8.5 cm}}
\caption{(Color online) Left: Lattice structure of the BLG. The unit cell contains atoms $a$ and $b$ on top layer and $a'$ and $b'$ on bottom layer. The intra-layer and inter-layer NN electron hoppings are $t$ and $t_1$, respectively. Right: First Brillouin zone and the two valleys $K$ and $K'$ in the momentum space.} 
\end{figure} 

Define the order parameters $m_j = (\langle\delta n_{j\uparrow}\rangle-\langle\delta n_{j\downarrow}\rangle)/2$ and $\rho_j = (\langle\delta n_{j\uparrow}\rangle+\langle\delta n_{j\downarrow}\rangle)$ for the spin and charge orderings, respectively. These parameters depend only on the index of the sublattice; within a sublattice, they are constants, $m_j = m_l$ and $\rho_j = \rho_l$, where the position $j$ belongs to the sublattice $l$. Because of the charge neutrality, we have $\rho_a = -\rho_{b'}$ and $\rho_b = -\rho_{a'}$, which comes from the broken layer-inversion symmetry. In terms of these order parameters, the average $\langle\delta n_{j\sigma}\rangle$ is given by $\langle\delta n_{j\sigma}\rangle = \sigma m_l + \rho_l/2$ where $j$ belongs to sublattice $l$ and $\sigma$ = 1 (-1) for spin up (down). The Hartree term in Eq. (\ref{ser}) can be written as  
\begin{eqnarray}
\Sigma^{\sigma H}(l,l)=-\sigma Um_{l} +(V_{ll}+U/2)\rho_l + V_{l\tilde l}\rho_{\tilde l},  \nonumber
\end{eqnarray}
where $\tilde l$ means that $\tilde a~(\tilde b) = b~(a)$ and $\tilde a'~(\tilde b') = b'~(a')$, and 
\begin{eqnarray}
V_{aa}&=& -v(r_{ab'}) + \sum_{\vec r\ne 0}[v(r)-v(|\vec r+\vec r_{ab'}|)],  \nonumber\\
V_{ab}&=&  \sum_{\vec r}[v(|\vec r+\vec r_{ab}|)-v(|\vec r+\vec r_{aa'})],  \nonumber\\
V_{bb}&=& -v(d)+\sum_{\vec r\ne 0}[v(r)-v(|\vec r-\vec d|)].  \nonumber
\end{eqnarray}
Here $v(r) = v_{ij}$ with $r$ the distance between the position $i$ and $j$, the $\vec r$-summations run over the positions on sublattice $a$, $\vec r_{ab'} = (1,1/\sqrt{3},-d)$ and $\vec r_{ab} = (1,1/2\sqrt{3},0)$ and $\vec r_{aa'} = (1,1/2\sqrt{3},-d)$ are respectively the vectors from atom $a$ to atoms $b'$, $b$ and $a'$ in the unit cell, and $\vec d = (0,0,d)$. The other quantities are given by $V_{a'a'} = V_{bb}$, $V_{b'b'} = V_{aa}$, and $V_{a'b'} = V_{b'a'} = V_{ab} = V_{ba}$.

In the exchange (XC) part, the average $\langle c_{i\sigma}c^{\dagger}_{j\sigma}\rangle$ can be a complex containing an imaginary part \cite{Varma},
\begin{eqnarray}
\langle c_{i\sigma}c^{\dagger}_{j\sigma}\rangle = R_{ij\sigma}+iI_{ij\sigma}. \label {av}
\end{eqnarray}
The imaginary part $I_{ij\sigma}$ corresponds to a current and is self-consistently determined by the approximation. In a recent work \cite{Yan4}, we have shown that within the range of physical interaction strength only the intra-sublattice current orderings are possible. There is no inter-sublattice current ordering because it breaks the translational invariance; more symmetry breaking would happen in a stronger interacting system. The remaining real part $R_{ij\sigma}$ for $i$ and $j$ in different sublattices gives rise to the renormalization of the inter-sublattice electron hoping. We suppose this renormalization has been already included in the original hoping terms. Therefore, we here consider only the current orderings (and the self-energies) between the sites of same sublattice. 

In momentum space, the exchange part of the self-energy is given by
\begin{eqnarray}
\Sigma^{\sigma XC}_l(k)= -\frac{1}{N}\sum_{k'} v^{eff}(|\vec k-\vec k'|)(\langle c^{\dagger}_{lk'\sigma}c_{lk'\sigma}\rangle-1/2),  \nonumber
\end{eqnarray}
where $N$ is the total number of the unit cells of the BLG lattice, $c^{\dagger}_{lk'\sigma}$ ($c_{lk'\sigma}$) creates (annihilates) an electron of momentum $k'$ and spin $\sigma$ on sublattice $l$, and $k'$-summation runs over the first Brillouin zone. Here, the main points are that (1) the quantity $\langle c^{\dagger}_{lk\sigma}c_{lk\sigma}\rangle-1/2$ as a function of $k$ is sizable only when $k$ is close to the Dirac points $K$ and $K'$ \cite{Yan2}, (2) for carrier concentration close to the charge neutral point, we need to consider only low-energy quasiparticles with $k$ close to the Dirac points, and (3) $v^{eff}(q)$ is a slowly varying function of $q$ because of the electron screening. Under these considerations, the exchange self-energy $\Sigma^{\sigma XC}_l(k)$ for $k$ in valley $v = K$ or $K'$ can be approximated as
\begin{eqnarray}
\Sigma^{v\sigma XC}_l&=& -\frac{1}{N}\sum_{v'k'} v^{eff}(|\vec v-\vec v'|)(\langle c^{\dagger}_{lv'+k'\sigma}c_{lv'+k'\sigma}\rangle-1/2)  \nonumber\\
&=& -\frac{v_c}{N}\sum_{v'k'} (\langle c^{\dagger}_{lv'+k'\sigma}c_{lv'+k'\sigma}\rangle-1/2) \nonumber\\
& &-s_v\frac{v_s}{N}\sum_{v'k'} s_{v'}\langle c^{\dagger}_{lv'+k'\sigma}c_{lv'+k'\sigma}\rangle \nonumber
\end{eqnarray}
where $k'$ is measured from the Dirac point $v'$ and the $k'$-summation runs over a circle $k' \leq 1/a$ in valley $v'$  [see Fig. 1 (right)], $v_{c,s} = [v^{eff}(0)\pm v^{eff}(2K)]/2$, and $s_v$ = 1 (-1) for $v = K$ ($K'$). The first term in the last equal can be written as $-v_c(\sigma m_l +\rho_l/2 +\delta/2)$ with $\delta$ as the average electron doping concentration per atom. The last term corresponds to the current ordering since the imaginary part in Eq. ({\ref{av}) is given by
\begin{eqnarray}
I_{ij\sigma}= \frac{1}{N}\sum_{vk} s_v\langle c^{\dagger}_{lv+k\sigma}c_{lv+k\sigma}\rangle \sin(\vec K\cdot\vec r_{ij}). \label{crt}
\end{eqnarray}
The `current' (up to a constant factor) $I_{ij\sigma}$ is finite only when the distributions in the two valleys are unbalanced. Since the sublattice is a triangular lattice, the current flows in three directions with equal magnitude. However, the current density at each atom vanishes. Note that the current $I_{ij\sigma}$ depends on the relative vector $\vec r_{ij}$ from position $i$ to $j$ and does not change the translational invariance of the system. Therefore, the current can exist in the uniform triangular lattice.

The total self-energy in momentum space $\Sigma^{v\sigma}_l = \Sigma^{\sigma H}(l,l) +\Sigma^{v\sigma XC}_l$ now can be written as
\begin{eqnarray}
\Sigma^{v\sigma}_{l} = \epsilon_l-\sigma u_0m_l -s_v\Delta_{l\sigma}-v_c\delta/2, \label{se}
\end{eqnarray}
where $\epsilon_l = u_{ll}\rho_l+u_{l\tilde l}\rho_{\tilde l}$ with $u_{ll} = V_{ll}+ U/2 -v_c/2$ and $u_{l\tilde l} = V_{l\tilde l}$, $u_0 = U+v_c$, and $\Delta_{l\sigma}$ is the current order parameter. The relation $\epsilon_l = -\epsilon_{\bar l}$ with $\bar a~(\bar b') = b'~(a)$ and $\bar b~(\bar a') = a'~(b)$ is valid because of the charge neutrality condition. Since the term $-v_c\delta/2$ is a constant (independent of the layer, valley, and spin), we hereafter will discard this term in the self-energy. The order parameters are calculated by
\begin{eqnarray}
\rho_l = \frac{1}{2N}\sum_{vk\sigma}(\langle c^{\dagger}_{lv+k\sigma}c_{lv+k\sigma}\rangle-\langle c^{\dagger}_{{\bar l}v+k\sigma}c_{{\bar l}v+k\sigma}\rangle), \label{o1}\\
m_{l}=\frac{1}{2N}{\sum_{vk\sigma}}\sigma\langle c^{\dagger}_{lv+k\sigma}c_{lv+k\sigma}\rangle, \label{o2}\\
\Delta_{l\sigma}=\frac{v_s}{N}{\sum_{vk}}s_v\langle c^{\dagger}_{lv+k\sigma}c_{lv+k\sigma}\rangle. \label{o3}
\end{eqnarray} 
The interaction parameters have been determined in the previous work \cite{Yan3} with the results: $u_{aa} \approx u_{bb} = 3.3 \epsilon_0$, $u_{ab} = 6.58 \epsilon_0$, $u_0 = 6.38\epsilon_0$, $v_c = 5.38\epsilon_0$, and $v_s = 6.372\epsilon_0$ with $\epsilon_0 = \sqrt{3}t/2$.

Define the operator
\begin{eqnarray}
C^{\dagger}_{vk\sigma}=(c^{\dagger}_{a,v+k,\sigma},c^{\dagger}_{b,v+k,\sigma}
,c^{\dagger}_{a',v+k,\sigma},c^{\dagger}_{b',v+k,\sigma}) \nonumber
\end{eqnarray}
The effective Hamiltonian under the MFT is obtained as
\begin{eqnarray}
H = \sum_{vk\sigma}C^{\dagger}_{vk\sigma}H_{vk\sigma}C_{vk\sigma}  \nonumber
\end{eqnarray}
with
\begin{eqnarray}
H_{vk\sigma} = \begin{pmatrix}
\Sigma^{v\sigma}_{1}& e_{vk}&0&0\\
e^{\ast}_{vk}&\Sigma^{v\sigma}_{2}&-t_1&0\\
0&-t_1&\Sigma^{v\sigma}_{3}&e_{vk}\\
0&0&e^{\ast}_{vk}&\Sigma^{v\sigma}_{4}\\
\end{pmatrix} \label{hmat}
\end{eqnarray}
where $e_{vk} = s_vk_x+ik_y$ in units of $\epsilon_0 = 1$, and the sublattice index $l$ runs from 1 to 4 for the sublattices $a, b, a'$, and $b'$, respectively. 

In the absence of an external magnetic field, we have shown that there is no spin ordering $m_l = 0$ \cite{Yan3}. Then, the current ordering parameters satisfy the relations, $\Delta_{1\sigma} = - \Delta_{4\sigma}$, $\Delta_{2\sigma} = - \Delta_{3\sigma}$, and $\Delta_{l\uparrow}=-\Delta_{l\downarrow}$ \cite{Yan3}. The charge ordering can appear only when an external voltage is applied between the two layers. With such a voltage, the electrons experience different potentials -$u$ and $u$ in the top and bottom layers, respectively. The Hamiltonian matrix $H_{vk\sigma}$ is then modified by adding to it a diagonal matrix 
\begin{eqnarray}
H_{ex} = {\rm Diag}\{-u,-u,u,u\}, \nonumber
\end{eqnarray}
or $\epsilon_1$ and $\epsilon_2$ in the self-energy are replaced with $\epsilon_1-u$ and $\epsilon_2-u$, respectively.

To proceed, we start with the Green's function of the electrons. The Green's function $G$ of the electron system in the imaginary $\tau$ space is defined as 
\begin{eqnarray}
G^{v\sigma}(k,\tau-\tau') = -\langle T_{\tau}C_{vk\sigma}(\tau)C^{\dagger}_{vk\sigma}(\tau')\rangle. \nonumber
\end{eqnarray}
In the Matsubara-frequency space, $G$ (a 4$\times$4 matrix) is expressed as
\begin{eqnarray}
G^{v\sigma}(k,i\omega_{\ell}) = (i\omega_{\ell} +\mu -H_{vk\sigma})^{-1} \label {Gk}
\end{eqnarray}
where $\mu$ is the chemical potential determined by
\begin{eqnarray}
\delta = \frac{1}{4N}\sum_{vk\sigma}[T\sum_{\ell}{\rm Tr}G^{v\sigma}(k,i\omega_{\ell})\exp(i\omega_{\ell}\eta)-2], \label {chm}
\end{eqnarray}
where $T$ is the temperature, and $\omega_{\ell} = (2\ell+1)\pi T$ is the Matsubara frequency, and $\eta$ is an infinitesimal small positive constant. 

Note that the Hamiltonian matrix can be transformed to a simple form. Denote the angle of the vector ($s_vk_x$, $k_y$) as $\phi_v$ and define the matrix
\begin{eqnarray}
M(\phi_v) = {\rm Diag}\{\exp(i\phi_v),1,1,\exp(-i\phi_v)\}.\nonumber
\end{eqnarray}
With $M(\phi_v)$, the transformed Hamiltonian $M^{\dagger}(\phi_v)H_{vk\sigma}M(\phi_v)\equiv h_{vk\sigma}$ is independent of the momentum angle. Similarly, we have $M^{\dagger}(\phi_v)G_{v\sigma}(k,i\omega_{\ell})M(\phi_v)\equiv g^{v\sigma}(k,i\omega_{\ell})$ independent of the angle $\phi_v$. It is then convenient to work in the space of the transformed Hamiltonian $h_{vk\sigma}$. By denoting the $\alpha$th component of the $\lambda$th eigenfunction of $h_{vk\sigma}$ with eigenvalue $E^{v\sigma}_{\lambda}(k)$ as $W^{v\sigma}_{\alpha\lambda}(k)$, the $\alpha\beta$th element of the Green's function $g^{v\sigma}$ is expressed as
\begin{eqnarray}
g^{v\sigma}_{\alpha\beta}(k,i\omega_{\ell}) &=& \sum_{\lambda}W^{v\sigma}_{\alpha\lambda}(k)W^{v\sigma}_{\beta\lambda}(k)/[i\omega_{\ell}+\mu-E^{v\sigma}_{\lambda}(k)]. \nonumber
\end{eqnarray}

For our purpose, we write the order parameters $\rho_l$ and $\Delta_{l\uparrow} \equiv \Delta_l$ in terms of the Green's function. Using the definition for the Green's function $g^{v\sigma}(k,i\omega_{\ell})$, we have
\begin{eqnarray}
\rho_l = \frac{1}{2N}\sum_{vk\sigma}[g^{v\sigma}_{ll}(k,i\omega_{\ell})-g^{v\sigma}_{\bar l\bar l}(k,i\omega_{\ell})], \label{od1}\\
\Delta_1 = \frac{v_sT}{N}\sum_{vk\ell}s_vg^{v\uparrow}_{11}(k,i\omega_{\ell}), \label {d1}\\
\Delta_2 = \frac{v_sT}{N}\sum_{vk\ell}s_vg^{v\uparrow}_{22}(k,i\omega_{\ell}). \label {d2}
\end{eqnarray}

\section{phase transition}

The phase boundary of the SPCS is the relation between the critical temperature $T_c$ and the carrier doping concentration $\delta$. We will consider the cases for zero and finite external voltages. 

\subsection{Zero Voltage}

For zero voltage, $u$ = 0, there is no charge ordering, $\rho_l$ = 0 and $\epsilon_l$ = 0 \cite{Yan3}. The Hamiltonian matrix $h_{vk\sigma}$ has the property $h_{vk\sigma}=Sh_{-vk\sigma}S = Sh_{vk-\sigma}S$, where $S = \tau_1\sigma_1$ with the Pauli matrix $\tau_1$ implying the exchange of top and bottom layers and $\sigma_1$ the exchange of $(a,b)$ and $(a',b')$ atoms. If $W^{v\sigma}(k)$ is an eigenfunction of $h_{vk\sigma}$ with eigenvalue $E^{v\sigma}$, then $SW^{v\sigma}(k)$ is an eigenfunction of $h_{-vk\sigma}$ or $h_{vk-\sigma}$ with the same eigenvalue. Therefore, the whole eigenstates can be obtained from the one only for a given spin in a single valley. Because of this property of the effective Hamiltonian, we only need to consider the Green's function in the $K$ valley for spin-up electrons. We hereafter drop the valley and spin subscripts $v$ and $\sigma$ in the Green's function and $g(k,i\omega_{\ell})$ is understood to be the Green's function in the $K$ valley for spin-up electrons. 

\begin{figure}[t]
\centerline{\epsfig{file=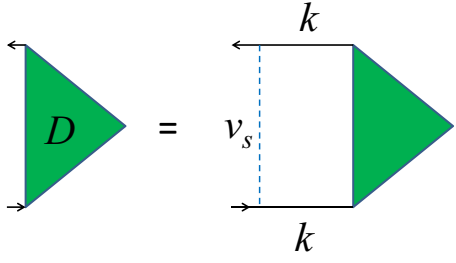,width=6.5 cm}}
\caption{(Color online) Diagrammatic equation for the matrix $D$ (green triangle). The solid lines are the Green's functions and the dashed line is the effective interaction $v_s$. } 
\end{figure} 

As we approach the phase boundary from the SPCS side, $\Delta_1$ and $\Delta_2$ become vanishingly small. We expand the equations (\ref{d1}) and (\ref{d2}) to the first order in $\Delta_1$ and obtain
\begin{eqnarray}
1 &=& -\frac{v_sT}{N}\sum_{k\ell}[(gDg)_{11}-(gDg)_{44}], \label {c1}\\
\frac{\partial\Delta_2}{\partial\Delta_1} &=& -\frac{v_sT}{N}\sum_{k\ell}[(gDg)_{22}-(gDg)_{33}], \label {c2} 
\end{eqnarray}
where $D = -\partial h_k/\partial\Delta_1$ is a matrix obtained as
\begin{eqnarray}
D = {\rm Diag}\{1,\frac{\partial\Delta_2}{\partial\Delta_1},-\frac{\partial\Delta_2}{\partial\Delta_1},-1\}.\nonumber
\end{eqnarray}
In deriving Eqs. (\ref{c1}) and (\ref{c2}), we have used $\partial g = -g(\partial g^{-1}) g = g(\partial h_k)g$. The Green's functions in Eqs. (\ref{c1}) and (\ref{c2}) are now calculated in the normal state with $\Delta_{1,2} = 0$. In the normal state, since the system is symmetric for the exchange of top and bottom layers, the term -$(gDg)_{44}$ in the sum of Eq. (\ref{c1}) gives rise to the same contribution as $(gDg)_{11}$. Similarly, the term -$(gDg)_{33}$ contributes the same as $(gDg)_{22}$ in Eq. (\ref{c2}). Therefore, the two summations in Eqs. (\ref{c1}) and (\ref{c2}) can be simplified. On the other hand, the summations over the Matsubara-frequency can be carried out immediately with the result given as
\begin{eqnarray}
T\sum_{\ell}g_{ll'}g_{l'l}&=& \sum_{\gamma\gamma'}W_{l\gamma}W_{l'\gamma}W_{l'\gamma'}W_{l\gamma'}F(E_{\gamma},E_{\gamma'})
\nonumber\\
&\equiv& f_{ll'}(k),\label{ff} 
\end{eqnarray}
and
\begin{eqnarray}
F(E_{\gamma},E_{\gamma'})=\frac{f(E_{\gamma})-f(E_{\gamma'})}{E_{\gamma}-E_{\gamma'}},\nonumber
\end{eqnarray}
where $f(E_{\gamma})$ is the Fermi distribution function, and $\gamma$ and $\gamma'$ run over the indexes of the four energy levels. When $E_{\gamma} = E_{\gamma'}$, $F$ is defined as $F = df(E_{\gamma})/dE_{\gamma}$. Now, Eqs. (\ref{c1}) and (\ref{c2}) can be rewritten in a compact form
\begin{eqnarray}
D_{l}= -\frac{2v_s}{N}\sum_{kl'}f_{ll'}(k)D_{l'}. \label{cc}
\end{eqnarray}
with $D_l$ the $l$th element in the diagonal of the matrix $D$. Recalling that $\Delta_l$'s represent the current orderings, the matrix $D$ actually describes the particle-hole propagator in the the current channel. The diagrammatic representation is shown in Fig. 2.

\begin{figure}[t]
\centerline{\epsfig{file=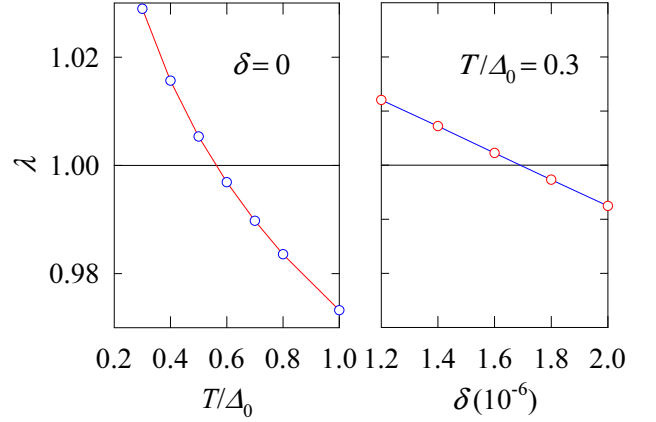,width=8.5 cm}}
\caption{(Color online) Left: Quantity $\lambda$ as a function of temperature $T$ at charge neutrality point $\delta = 0$. Right: $\lambda$ as function of $\delta$ at $T/\Delta_0$ = 0.3.} 
\end{figure} 

To search the phase boundary, we need to solve the Green's function at a series of selected points ($\delta, T$) in the normal phase. For a given carrier concentration $\delta$, the transition temperature $T_c$ is found by gradually lowering temperature $T$ from a value higher than $T_c$. At each point ($\delta, T$), we self-consistently solve Eq. (\ref{cc}) for $l$ = 2 to determine $\partial\Delta_2/\partial\Delta_1$. Then we apply the result in the right-hand side of Eq. (\ref{cc}) for $l$ = 1 and denote the calculated value as $\lambda$. By inspecting this value $\lambda$, the transition temperature $T_c$ is reached when $\lambda$ is unity. Figure 3 (left) shows the value $\lambda$ as a function of temperature $T$ at charge neutrality point $\delta = 0$. The transition temperature $T_c$ at $\delta = 0$ is determined as $T_c = 0.567\Delta_0$ with $\Delta_0$ = 1 meV the gap parameter observed by experiment \cite{Velasco}. However, at doping concentration $\delta > 1.4\times10^{-6}$, the transition temperature is not a one-to-one correspondence with the doping. In this case, we solve the equation with varying doping at a fixed temperature. In Fig. 3 (right), $\lambda$ is presented as a function of $\delta$ at $T/\Delta_0$ = 0.3. We have thus determined the phase boundary of the SPCS. The result is shown in Fig. 4. The highest $T_c = 0.567\Delta_0$ appears at the CNP $\delta = 0$. The largest carrier concentration for the SPCS is about $\delta \approx$ 1.7$\times 10^{-6}$ with $T_c/\Delta_0 \approx$ 0.3. Note that the Fermi energy is $E_F = 8\pi\delta\epsilon^2_0/\sqrt{3}t_1$. At $\delta$ = 1.7$\times 10^{-6}$, we have $E_F/\Delta_0$ = 0.61. Therefore, the Fermi energy at the largest carrier concentration for the SPCS is about the same order of magnitude as the largest $T_c$ at the CNP. 

Since the Hamiltonian is symmetric about the carrier doping, $T_c$ is an even function of $\delta$. 

\begin{figure}[t]
\centerline{\epsfig{file=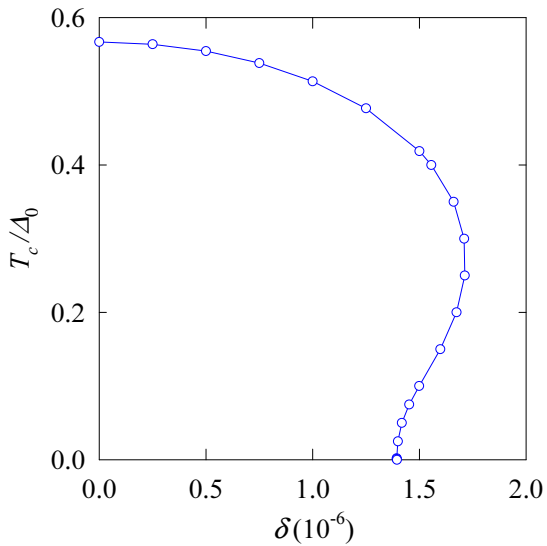,width=7.5 cm}}
\caption{(Color online) Phase boundary of the spin-polarized current state.} 
\end{figure} 

\subsection{Finite Voltage}

At finite voltage, the system is layer-charge polarized with $\rho_l \neq 0$. In Fig. 5, we present the charge order parameters $\rho_1$ and $\rho_2$ of electrons as the functions of the external potential $u$ at the charge neutrality point $\delta = 0$. The temperature is at the transition point $T = 0.567\Delta_0$ for $u$ = 0. The potential difference between the bottom layer and top layer is $2u$. For positive $u$, the polarized electron number per unit cell at top (bottom) layer is $\rho_1+\rho_2 > 0$ ($\rho_3+\rho_4 = -\rho_2-\rho_1< 0$). The polarization increases with increasing $u$. 

\begin{figure}[b]
\centerline{\epsfig{file=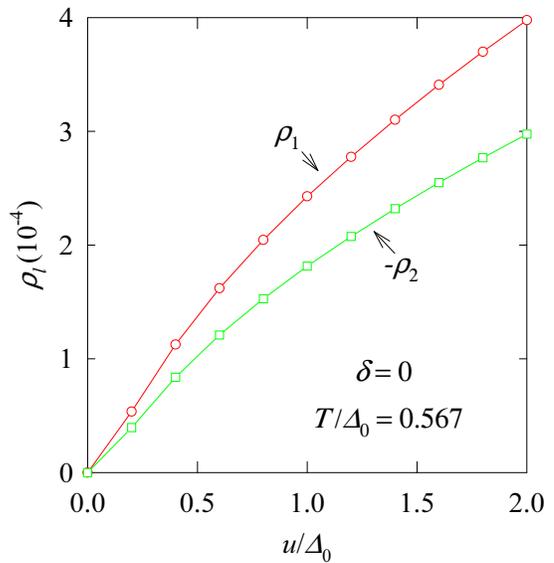,width=7.5 cm}}
\caption{(Color online) Charge order parameters $\rho_1$ and $\rho_2$ as functions of the external potential $u$ at $\delta = 0$ and $T/\Delta_0 = 0.567$.} 
\end{figure} 

At low temperature and low carrier concentration, the current ordering may coexist with the charge ordering when $u \neq 0$. To search the boundary of the spin-polarized current phase, we take the derivative of the order parameters with respect to $\Delta_1$. From Eqs. (\ref{d1}) and (\ref{d2}), we have
\begin{eqnarray}
1 &=& -\frac{v_sT}{N}\sum_{vk\ell}[g^{v\uparrow}(k,i\omega_{\ell})Dg^{v\uparrow}(k,i\omega_{\ell})]_{11}, \label {cf1}\\
\frac{\partial\Delta_2}{\partial\Delta_1} &=& -\frac{v_sT}{N}\sum_{vk\ell}[g^{v\uparrow}(k,i\omega_{\ell})Dg^{v\uparrow}(k,i\omega_{\ell})]_{22}. \label {cf2} 
\end{eqnarray}
Since the layer inversion symmetry is now broken, these equations are different from Eqs. (\ref{c1}) and (\ref{c2}). Note that the dependence of the charge ordering $\rho_l$ on $\Delta_1$ is negligible small since $\rho_l$ is mainly determined by the external voltage. (We have numerically checked this point.) The summations over the Matsubara frequency in Eqs. (\ref{cf1}) and (\ref{cf2}) can be performed similarly as shown in Eq. (\ref{ff}). The phase boundary of the SPCS is now determined by Eqs. (\ref{cf1}) and (\ref{cf2}) with $\Delta_l = 0$ in the Green's function. 

The obtained phase boundary of the SPCS at finite $u$ is shown in Fig. 6. By comparing the case of zero $u$ shown in Fig. 4, the phase area of the SPCS shrinks with increasing $u$. The phase of the SPCS eventually disappears at certain strength of the potential difference $u$. As seen from Fig. 6, the unusual feature of the phase diagram for a finite $u$ in certain range of strength is that there are two transition temperatures for a given carrier concentration. We analyze this result below. 

\begin{figure}[t]
\centerline{\epsfig{file=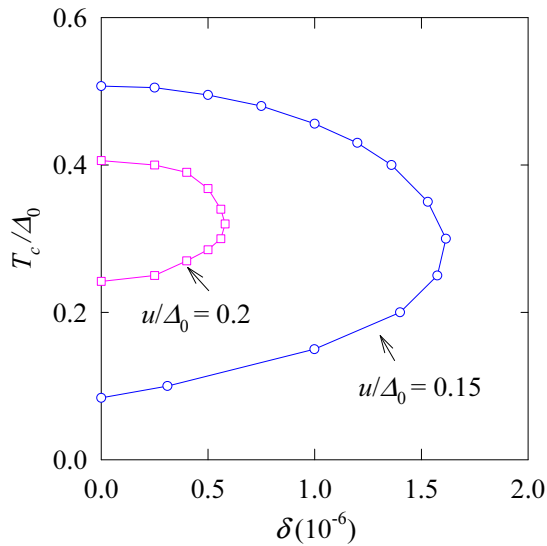,width=7.5 cm}}
\caption{(Color online) Phase boundary of the spin-polarized current state at finite potential difference $u$ between bottom and top layer.} 
\end{figure} 

First, there is a gap between the conduction and valence bands because of the finite potential $u$. At low temperature close to zero, for carrier concentration close to the CNP, the chemical potential $\mu$ (approximately the Fermi energy) is close to the bottom of the conduction band. The current ordering happens when there exist a valley polarization because of the exchange effect; the energy levels of spin-$\sigma$ electrons in one valley are raised with $\Delta_{1\sigma}$ while the levels are lowered by -$\Delta_{1\sigma}$ in another valley, resulting in the spin-$\sigma$ electrons transferring from the former to the latter valley. The level change $\Delta_{1\sigma}$ and the electron transferring are self-consistently determined by themselves. Below the first transition temperature, this process cannot happen because there are not enough electrons below the level $\mu$ in the conduction band for transferring. However, with increasing the temperature, the electrons in the valence band can be excited to the conduction band. Especially, above the first transition temperature, the excited electrons can participate in the transferring process and assist the current ordering. On the other hand, the thermal excitations of electrons between two valleys are also allowable and are weakening the exchange effect. At higher temperature above the second transition temperature, the exchange effect is quenched by the thermal excitations and there is no current ordering. Therefore, there is a second transition temperature higher than the first one.

In Fig. 4, we have seen that there are two transition temperatures for $1.4\times10^{-6} < \delta < 1.7\times10^{-6}$ where the external voltage is zero. Within this doping range and below the first transition temperature, there is no gap between the conduction and valence bands. The SPCS emerges above the first transition temperature just because the thermal excitations of electrons from the low levels in one valley to the levels above the chemical potential in another valley assist the electron transferring from the former to the latter valley. The mechanism for the two transition temperatures is the same as explained above. 

\section{summary}

Using the four-band model, we have studied the phase boundary of the spin-polarized current state of the interacting electrons in bilayer graphene. In the absence of external voltage, the highest transition temperature is found as $T_c = 0.567\Delta_0$ = 0.567 meV appearing at the charge neutrality point $\delta = 0$. The SPCS phase extends to a carrier concentration about $\delta \approx$ 1.7$\times 10^{-6}$ with $T_c \approx$ 0.3 meV. At finite voltage between the two layers, we find there are two transition temperatures corresponding to a given carrier concentration. The physics of such an unusual phase boundary is explained as the two effects of the thermal excitations: (1) the excited electrons participate in the process of transferring from one valley to another valley and assist the current ordering, and (2) excitations between two valleys at higher temperature quench the current ordering. 

The result should be useful for real application of the BLG.

\acknowledgments

This work was supported by the National Basic Research 973 Program of China under Grant No. 2012CB932302 and the Robert A. Welch Foundation under Grant No. E-1146.

\end{document}